\documentclass[sigconf, 10pt, nonacm, screen]{acmart}
\newcommand{\secref}[1]{\S{}\ref{#1}}
\newcommand{\figref}[1]{Fig.~\ref{#1}}

\newcommand{\focusapp}[0]{HPDT}
\newcommand{\focusapps}[0]{HPDTs}
\newcommand{\afocusapp}[0]{an HPDT}

\newcommand{\arch}[0]{DBPT}

\newcommand{\us}[0]{\textmu{}s}

\newcommand{\smpc}[0]{MPC}

\newcommand{\LANWithoutExplanation}{mohassel2017secureml, liu2017minionn, juvekar2018gazelle, chandran2019ezpc, riazi2019xonn, mishra2020delphi, watson2022piranha, ren2024accelerating}

\title{Fast Networks for High-Performance Distributed Trust}

\author{Yicheng Liu}
\affiliation{%
  \institution{UCLA}%
  \country{}
}
\email{easonliu@cs.ucla.edu}

\author{Rafail Ostrovsky}
\affiliation{%
  \institution{UCLA}%
  \country{}
}
\email{rafail@cs.ucla.edu}

\author{Scott Shenker}
\affiliation{%
  \institution{UC Berkeley and ICSI}%
  \country{}
}
\email{shenker@icsi.berkeley.edu}

\author{Sam Kumar}
\affiliation{%
  \institution{UCLA}%
  \country{}
}
\email{samkumar@cs.ucla.edu}

\begin{abstract}
Organizations increasingly need to collaborate by performing a computation on their combined dataset, while keeping their data hidden from each other.
Certain kinds of collaboration, such as collaborative data analytics and AI, require a level of performance beyond what current cryptographic techniques for distributed trust can provide.
This is because the organizations run software in \emph{different trust domains}, which can require them to communicate over WANs or the public Internet.
In this paper, we explore how to instead run such applications using fast datacenter-type LANs.
We show that, by carefully redesigning distributed trust frameworks for LANs, we can achieve up to order-of-magnitude better performance than na\"ively using a LAN.
Then, we develop deployment models for \emph{Distributed But Proximate Trust (DBPT)} that allow parties to use a LAN while remaining physically and logically distinct.
These developments make secure collaborative data analytics and AI significantly more practical and set new research directions for developing systems and cryptographic theory for high-performance distributed trust.
\end{abstract}

\begin{document}

\maketitle

\section{Introduction}

\emph{Distributed trust} is a technique for building efficient systems with cryptographic security properties.
With distributed trust, $n$ different parties work together to run a system, and the system's security guarantees hold if up to $m$ of those parties (for some $m$, $n$ such that $1 \leq m \leq n - 1$) are compromised by an adversary.
It enables applications like anonymous communication~\cite{chaum1981mixnets, corrigangibbs2015riposte, kwon2020xrd}, metadata-hiding file sharing~\cite{chen2020metal, chen2022titanium}, permissioned blockchains~\cite{androulaki2018hyperledger}, private blocklists~\cite{kogan2021checklist}, and privacy-preserving statistics collection~\cite{corrigangibbs2017prio, boneh2021poplar, addanki2022prioplus}.

Certain promising distributed trust applications are not widely deployed because they require a level of performance that is beyond what current techniques can provide.
We refer to them as \emph{high-performance distributed trust} applications, or \focusapps{}, and they are the focus of this paper.
One example of \afocusapp{} is collaborative data analytics~\cite{bater2017smcql, volgushev2019conclave,falk2019durasift,ion2020deploying}, in which multiple organizations perform data analysis on their combined dataset, while keeping their data hidden from each other.
Other examples are AI model training on a combined dataset~\cite{rouhani2018deepsecure, watson2022piranha} and privacy-preserving AI inference~\cite{mohassel2017secureml, juvekar2018gazelle, riazi2019xonn, mishra2020delphi, rathee2020cryptflow2, lehmkuhl2021muse}.
\focusapps{} have drawn industry interest, including from the financial~\cite{eizenman2019scotiabank} and healthcare~\cite{owkin2022federatedlearning} sectors, and have incipient adoption such as in Meta's Private Lift~\cite{reyes2022privatelift} and Google's Private Join and Compute~\cite{walker2019privatejoinandcompute}.
Still, \focusapp{} adoption remains limited and isolated, due to performance barriers seemingly inherent in distributed trust.

The performance cost of distributed trust stems not only from its cryptography, but also from its deployment model.
In a distributed-trust system, each party must be deployed in a \emph{different trust domain}, with no central point of attack or single administrative party spanning multiple trust domains~\cite{dauterman2022reflections}.
In distributed trust deployments in industry, each party often hosts their component of the system in their own computing infrastructure (or \emph{internal cloud})~\cite{polychroniadou2023primematch, verma2020odoh}.
If a distributed trust application is run in a public cloud, it is often seen as a requirement that parties are in \emph{different} clouds (e.g., one party in Amazon AWS and another party in Microsoft Azure) so that no single cloud provider becomes a central point of attack~\cite{lund2019svr, kumar2021mage, geoghegan2022exposure, lindell2023mpc, kaviani2024flock, connell2024svr3}, or at the very least in different regions of the same cloud~\cite{wang2017agmpc, corrigangibbs2017prio, chen2020metal, mishra2020delphi, lehmkuhl2021muse}.
Thus, the $n$ parties may have to communicate using \textbf{wide-area networks (WANs) or the public Internet}.

This is costly~\cite{lindell2023mpc} because the underlying cryptography often requires the $n$ parties to exchange very large amounts of data or exchange data over many round trips, proportional to the size of the computation.
The cost is not only in performance, but also monetary, as large data transfers incur high egress costs in the cloud~\cite{jain2023skyplane, prince2021egress}.
For \focusapps{}, these costs are compounded by increasing sizes of datasets and AI models.

Today, our community assumes that, for distributed trust, parties may have to communicate via WANs or the public Internet.
Researchers routinely measure empirical performance with parties in distinct geographic regions or cloud regions (or with networks that emulate this)~\cite{doerner2017floram, wang2017agmpc, chen2020metal, mishra2020delphi, rathee2020cryptflow2, lehmkuhl2021muse, kumar2021mage, zheng2021cerebro, dauterman2022waldo, watson2022piranha, vadapalli2023duoram}.
Communication and round complexity are defining performance metrics in cryptographic protocol design~\cite{chen2017psi, rathee2020cryptflow2, doerner2017floram, vadapalli2023duoram, falk2023gigdoram}, and support for geographically distant parties is seen as a hallmark of scalability~\cite{wang2017agmpc}.

Certain \focusapp{} systems use security models in which one trusted administrator~\cite{sharemind, blockdaemon, alchemy} or cloud provider~\cite{falk2023gigdoram} controls all parties' infrastructures, e.g., in one datacenter.
Systems researchers/practitioners view this as weaker than true distributed trust~\cite{dauterman2022reflections, kaviani2024flock, lindell2023mpc}, and we share this view.
For example, the parties could just have the administrator run the entire application on their behalf (e.g., via a clean room~\cite{awscleanrooms, gcpcleanrooms, snowflakecleanrooms, databrickscleanrooms}), without cryptographic \focusapp{} protocols.
We suspect that such weaker security models have arisen because \focusapps{} perform far better over LANs than over WANs.
Indeed, security researchers routinely measure \focusapp{} protocols over LANs as a point of comparison~\cite{\LANWithoutExplanation}, but without discussing implications for distributed trust.
The missing piece is an architecture for deploying \focusapps{} in datacenter-type LANs while retaining true distributed trust.

This paper aims to supply that missing piece.
We begin by showing that designing \focusapp{} software specifically for fast LANs is a compelling research direction that can yield performance gains of up to an order of magnitude in widely used frameworks like EMP-Toolkit~\cite{emp-toolkit} and MP-SPDZ~\cite{keller2020mpspdz} (\secref{sec:performance}).
These performance gains directly apply to systems that use weaker security discussed above (\secref{sec:centralized}).
To realize these performance gains \emph{without} sacrificing distributed trust, we develop architectures for \textbf{Distributed But Proximate Trust (\arch{})} that enable widespread deployment of \focusapps{} using fast LANs (\secref{sec:model}).
We then outline research directions to further accelerate \focusapps{} in such architectures (\secref{sec:future}).

Our message to the community is to: (1) treat fast LANs as a primary target for deploying \focusapps{}, (2) further develop \arch{} architectures for deploying \focusapps{} in LANs, and (3) design and develop \focusapp{} software to fully leverage fast networks.
We hope that this agenda will remove performance obstacles that currently gate widespread adoption of \focusapps{}.
\section{Background}
\label{sec:background}

\subsection{Secure Multi-Party Computation}
\label{s:smpc}

The full generality of distributed trust is captured by a family of cryptographic tools called Secure Multi-Party Computation (\smpc{})~\cite{yao1982protocols, goldreich1987mentalgame, benor1988bgw}.
\smpc{} enables $n$ parties, each with secret data $x_1, \ldots, x_n$, to compute $f(x_1, \ldots, x_n)$, for any function $f$ of their choice, while ensuring that no party learns anything about any other party's secret data except for what is inherently revealed by the output of the function $f$.

In \secref{sec:performance}, we focus on \emph{generic} \smpc{} protocols capable of working with \emph{any} function $f$.
Generic \smpc{} protocols work by expressing $f$ as a boolean or arithmetic circuit $C$, where wires represent encrypted or secret-shared~\cite{shamir1979share} values and gates represent operations on these values.
These protocols are network-intensive because \emph{communication among the parties is proportional to the size of $C$}.
We next describe two types of generic \smpc{}, focusing on how they use the network.

\subsubsection{Garbled Circuits}

In \smpc{} protocols based on \emph{garbled circuits}~\cite{yao1982protocols, malkhi2004fairplay, buescher2015parallelization, yakoubov2017gentle, bicer2017yao, emp-toolkit}, wires represent encrypted bits, and gates represent AND and XOR operations over those bits.
Classically, garbled circuits work with $n = 2$ parties, called \emph{garbler} and \emph{evaluator}, but generalizations to more than 2 parties exist~\cite{beaver1990bmr, wang2017agmpc, goyal2025roundcollapsing, heath2025multiparty}; in any case, all parties execute the circuit.
Executing an XOR gate requires no communication and is concretely
very fast~\cite{kolesnikov2008freexor}.
Executing an AND gate (over two encrypted \emph{bits}), however, requires the garbler to send 32 bytes of data (two ciphertexts of 128 bits each) to the evaluator~\cite{zahur2015halfgates}.
Data transfer is in one direction (i.e., one network round) and the data can be streamed concurrently with circuit execution~\cite{huang2011faster}.
Thus, garbled-circuit-based \smpc{} is \emph{bandwidth-intensive} in network use.

\subsubsection{Secret Sharing}

In \smpc{} protocols based on \emph{secret sharing}~\cite{goldreich1987mentalgame, damgaard2012spdz, damgaard2013practical, keller2020mpspdz}, wires represent secret-shared~\cite{shamir1979share} values over an algebraic structure (e.g., integers modulo a prime), and gates represent addition and multiplication operations over those values.
Executing an addition gate requires no communication.
Executing a multiplication gate requires the parties to exchange data.
The network round must complete before the output of the multiplication gate is obtained, so the total number of network rounds is the depth of the circuit in multiplication gates.
Thus, secret-sharing-based \smpc{} is \emph{round-intensive} in how it uses the network.

\subsection{Network Costs of \smpc{}}

\focusapps{} are often deployed with parties communicating over WANs or the public Internet.
Such networks have high latency (e.g., milliseconds to 100s of milliseconds).
This limits performance for secret-sharing-based \smpc{}, for which each network round experiences a full RTT of latency.
Additionally, TCP often achieves only limited bandwidth over such networks.
This degrades performance for garbled-circuit-based \smpc{}, which can become bandwidth-bound.

In principle, it is possible to sidestep bandwidth-related performance issues, since large amounts of bandwidth \emph{exist} over the wide area.
For example, one can use multiple parallel TCP connections to obtain higher bandwidth for \smpc{}~\cite{kumar2021mage}.
Alternatively, one could statically allocate bandwidth for \smpc{} over a private WAN (e.g., prioritizing \smpc{} traffic over other traffic sources).
Unfortunately, such techniques have a significant monetary cost.
Large data transfers incur high data egress costs in the cloud~\cite{jain2023skyplane, prince2021egress}, and purchasing large amounts of wide-area network bandwidth is expensive.
Fundamentally, it is expensive to build and deploy high-bandwidth WANs~\cite{hong2018b4after}, and the costs are passed on to users.

\subsection{Alternatives to \smpc{}}

We briefly discuss two alternative approaches to enabling \focusapps{}, to explain why \smpc{} is preferable in certain cases.

\subsubsection{Fully Homomorphic Encryption (FHE)}
\label{s:fhe}

FHE allows general computation on encrypted data.
Given the encryptions of $n$ items $x_1, \ldots, x_n$, FHE allows one to compute the encryption of $f(x_1, \ldots, x_n)$ for any function $f$, without knowing the secret key.
Unlike \smpc{}, computing $f$ using FHE does not require communication among parties, except for providing the input at the start and decrypting the output at the end.

Unfortunately, FHE usually cannot be used as a replacement for \smpc{}.
The reason is that FHE is \emph{three to four (or more) orders of magnitude more computationally expensive} than \smpc{}.
Whereas the CPU time to execute \smpc{} gates is measured in nanoseconds or microseconds, FHE operation times are measured in milliseconds~\cite{viand2021fhecompilersok, kumar2023rethinking}.
For example, the ``bootstrap'' FHE operation, necessary for executing circuits with high multiplicative depth, takes several milliseconds to complete at best~\cite{li2024bootstrapping}.
That said, in settings where network communication is very limited or expensive, FHE may still be preferable to \smpc{}, despite its computational overhead, because it uses the network much more sparingly.

\subsubsection{Trusted Execution Environments (TEEs)}
\label{s:tees}

TEEs, like Intel SGX/TDX and AMD SEV-SNP, are a hardware feature that allows software to run in an environment isolated even from kernel and hypervisor software~\cite{baumann2014haven}.
They work by cryptographically securing the TEE's data while it is in memory, decrypting data as they enter the CPU caches.
Their security assumes a trusted hardware design that must not expose TEEs' secret keys to any software, and a silicon root of trust for attesting that the intended code is running in the TEE.

TEEs have been proposed as an alternative to \smpc{} for \focusapps{}~\cite{zheng2020sharing}.
The idea is to collect all parties' data within a TEE, allowing computation on all parties' data to take place within the TEE.
The TEE is hosted at a single party, but the TEE's security guarantees prevent that party from directly observing or influencing the data inside the TEE.

TEEs, however, have a major downside compared to \smpc{}: They have been found, time and time again during their short history, to be susceptible to security attacks, particularly via side channels~\cite{vanbulck2018foreshadow, sgaxe, lee2020membuster, murdock2020plundervolt, li2021cipherleaks, borrello2022aepicleak, vanschaik2024sgxfail}.
Such attacks potentially allow the party hosting the TEE to bypass its protections and access data inside the TEE, compromising the other parties' data.
Given the complexity of modern CPUs, it may not be possible to make them totally side-channel free.
\section{\smpc{} System Design for Fast LANs}
\label{sec:performance}

Researchers measure performance of their distributed-trust systems not only in WANs, but also in LANs, usually as a point of comparison~\cite{demmler2015aby, wang2017agmpc, keller2018overdrive, kumar2021mage, watson2022piranha, falk2023gigdoram, ren2024accelerating}.
Depending on the nature of the application and WAN setup, the performance gain from using a LAN instead of a WAN has been shown to be significant---for example, up to nearly $10\times$~\cite{watson2022piranha}, with increasing speedups for $n > 2$ parties~\cite{wang2017agmpc}.
RDMA has been shown to provide $1.2$--$4.2\times$ additional performance~\cite{ren2024accelerating}.

However, there is a difference between just running \focusapp{} software in a LAN, and properly designing \focusapp{} software to fully benefit from LANs.
In this section, we initiate the study of how to build \focusapp{} systems that can fully leverage fast LANs.
We present initial work that shows an \emph{additional} $10\times$  improvement via lightly redesigned \smpc{} frameworks.

As an analogy, consider how the emergence of kernel-bypass networking initiated a rich line of research on how to best use it in regular (i.e., non-distributed-trust) datacenter systems and applications~\cite{kalia2016fasst, kalia2019erpc, wei2020xstore, kim2021linefs, schuh2021xenic, amaro2020rmc, burke2021prism, wei2018drtmh}.
Fast networks for \focusapps{} (e.g., enabled by DBPT architectures in \secref{sec:model}) similarly raise the bar for \focusapp{} systems.

\subsection{Experimental Methodology}
\label{s:methodology}

We use two single-socket servers.
Each server has an Intel Xeon Gold 5520+ Emerald Rapids CPU, a 10 Gbps Intel X710-T2L NIC (X710), and a 200 Gbps Nvidia MCX755106ASA-HEAT ConnectX-7 NIC (CX-7).
The X710 NICs are connected via an RJ45 cable, and the CX-7 NICs are connected via a QSFP56 Direct Attach Copper cable.
In both cases, the NICs are directly connected, with no intervening switches.

As a baseline for \focusapps{} running over a LAN representative of prior work, we use the Linux network stack with the X710 NICs.
For the X710 NICs, \texttt{ping} reports an RTT of $\approx$~800~\us{}.
While it is possible to obtain more than 10 Gbps locally or over the public Internet, the 10 Gbps provided by X710 is sufficient for our purposes; a modern garbled circuit protocol~\cite{zahur2015halfgates, emp-toolkit} (bandwidth-intensive type of \smpc{}) with one CPU core per party does not saturate a 10 Gbps link~\cite{kumar2021mage}.

For a setup optimized for LAN environments, we use RDMA with the CX-7 NICs.
For the CX-7 NICs, \texttt{ib\_send\_lat} reports a typical one-way (e.g., half-RTT) latency of $\approx$~1~\us{} for two-sided RDMA, and \texttt{ib\_read\_lat} and \texttt{ib\_write\_lat} report typical one-way latencies (e.g., half-RTTs) of $\approx$~2~\us{} and $\approx$~1~\us{}, respectively, for one-sided RDMA.

For each type of \smpc{}, for the function $f$ to run, we use multiplication of a $1024 \times 1024$ matrix with a $1024$-element vector, where one party provides the matrix and the other party provides the vector.
We measure baseline performance with the X710 NICs, representative of LAN experiments in prior work.
Then, we switch to the CX-7 NICs and iteratively redesign the software to better benefit from the LAN setting.

\subsection{Garbled-Circuit-Based \smpc{}}
\label{s:garbled_circuit_experiments}

\begin{figure}[t]
    \centering
    \includegraphics[width=0.95\linewidth,trim={0, 0.7cm, 0, 0.6cm},clip]{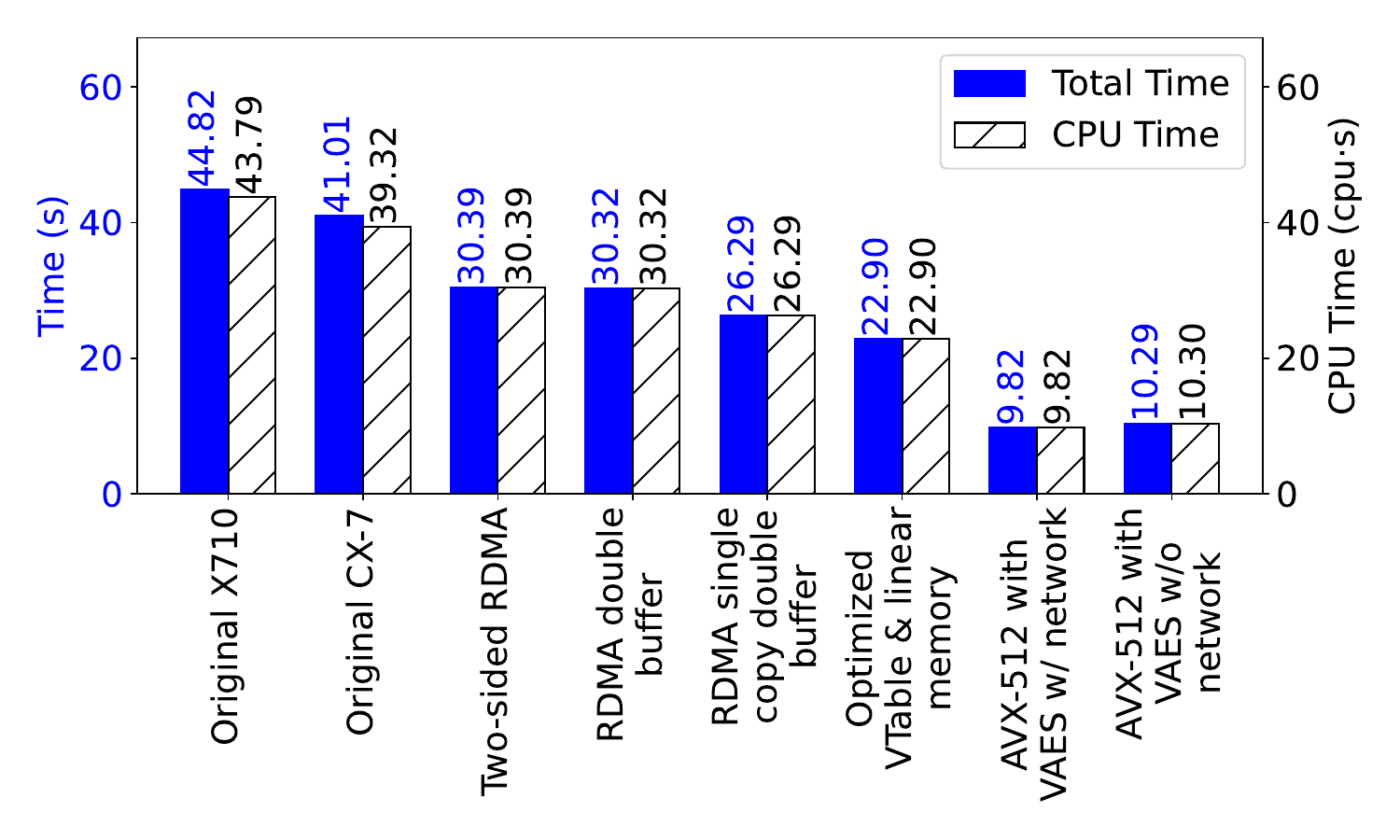}
    \caption{Total time and CPU time for EMP-Toolkit.}
    \label{fig:emptoolkit}
\end{figure}

We study garbled-circuit-based \smpc{} using the widely used EMP-Toolkit framework~\cite{emp-toolkit}, in a semi-honest threat model.
We run our workload with the matrix and vector containing 32-bit integers.
\figref{fig:emptoolkit} shows the wall-clock time and CPU time for each setup.
Each setup builds on the one to its left.

As shown in \figref{fig:emptoolkit}, simply upgrading from the commonly used 10 Gbps (X710) NIC to a 200 Gbps (CX-7) NIC does not significantly improve performance.
This is because the program is not bottlenecked by available network bandwidth.
However, using RDMA via \verb|ibverbs|, instead of the Linux network stack, significantly improves performance.
This is because RDMA eliminates the CPU overhead of the Linux network stack, allowing more CPU time to be spent on cryptographic work of the garbled circuit protocol.
Optimizations to use RDMA more efficiently, such as double buffering and single-copy, further enhance performance.

Once we use RDMA to eliminate Linux network stack overheads, other software overheads that were previously overshadowed by the cost of using the network are now significant.
For example, for software extensibility purposes, EMP-Toolkit implements AND and XOR gates as virtual function calls.
Eliminating these virtual function calls in favor of template-based static polymorphism, together with using an inlined memory layout, results in a noticeable speedup.

Similarly, AND gates require frequent hash operations~\cite{guo2020mitccrh}, primarily based on AES.
To improve performance, we changed EMP-Toolkit from using AES-NI instructions~\cite{bellare2013fixedkey} to using VAES instructions.
Fully utilizing VAES hardware required a new vectored/batched abstraction for programming $f$.
In total, the techniques bring $\approx$~$4.5\times$ speedup over the original.

Finally, we completely remove use of the network (which affects correctness) and observe that, even then, performance does not improve.
This suggests that further optimizing how we use RDMA is unlikely to yield more performance gains.

\subsection{Secret-Sharing-Based \smpc{}}
\label{s:secret_sharing_experiments}

\begin{figure}
    \centering
    \includegraphics[width=0.95\linewidth, trim={0, 0.7cm, 0, 0.6cm},clip]{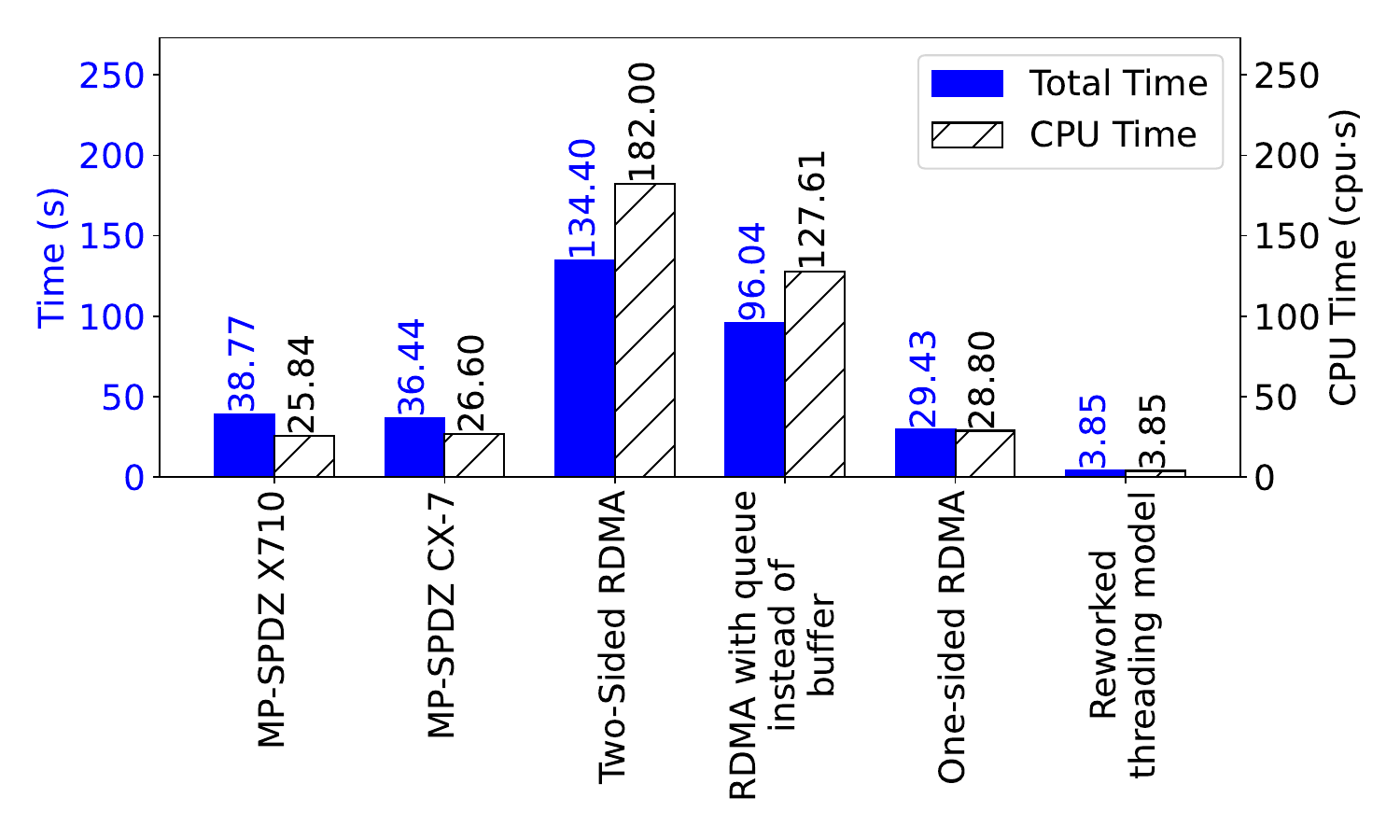}
    \caption{Total time and CPU time for MP-SPDZ.}
    \label{fig:mpspdz}
\end{figure}

We now study secret-sharing \smpc{} using the \mbox{``$\!\!\!\!\mod{2^k}$''} algorithm in the MP-SPDZ framework~\cite{keller2020mpspdz}, in a semi-honest threat model.
We run our workload with the matrix and vector containing 64-bit integers ($k = 64$); in MP-SPDZ, it incurs many network rounds.
\figref{fig:mpspdz} shows the wall-clock time and CPU time, including both online and preprocessing phases.

``Two-sided RDMA'' in \figref{fig:mpspdz} includes the RDMA-related optimizations in \secref{s:garbled_circuit_experiments}.
However, its performance is worse than using the Linux network stack.
This is because, whereas garbled circuits are \emph{bandwidth-intensive}, secret-sharing \smpc{} is \emph{round-intensive} (\secref{s:smpc}).
As a result, our design focus must shift from maximizing throughput to minimizing latency.
Accordingly, we replaced the RDMA buffering system from \secref{s:garbled_circuit_experiments} with a queue-based system: Each new piece of data triggers a new RDMA send request, which is matched with a receive request at the recipient.
We also switched from using two-sided RDMA to using one-sided RDMA, reducing latency and CPU overhead associated with synchronization.

MP-SPDZ originally used a threading model that assumed that network latency is much higher than the cost of thread switching and synchronization.
But, in our optimized LAN setting, this assumption no longer holds.
Thus, we redesigned the threading model, switching from a multi-threaded execution model to a single-threaded execution model.

Overall, these techniques achieve $\approx$~$10\times$ speedup---an order of magnitude---over the original version, and $\approx$~$40\times$ speedup compared to the first RDMA version.

\subsection{Discussion and Future Opportunities}
\label{s:future_performance_opportunities}

In our efforts above, just using RDMA instead of the Linux network stack provides a marginal improvement, at best.
We believe the reason is that existing \smpc{} frameworks were designed assuming slower networks.
Once we applied kernel bypass to fix network bottlenecks, new bottlenecks emerged.
Thus, RDMA serves as a hardware foundation for optimization, enabling subsequent software design changes to yield up to an order-of-magnitude performance improvement.

An opportunity to further accelerate \smpc{} over fast LANs is parallel execution using multiple CPU cores~\cite{buescher2015parallelization}.
EMP-Toolkit and MP-SPDZ, despite being widely used, do not parallelize circuit execution across multiple cores.
This may be because they were designed assuming the network is the bottleneck, so that using multiple cores would not help.
This is no longer the case over a fast LAN.
In our experiments above, with a single core per party, our optimized EMP-Toolkit uses $\approx$~30 Gbps of unidirectional bandwidth and our optimized MP-SPDZ uses $\approx$~5 Gbps of bidirectional bandwidth, out of 200~Gbps provided by the NIC.
This suggests that parallel execution is a promising performance opportunity.

Another opportunity is to design the machine shape.
For example, the $\approx$~30 Gbps required by our optimized single-core garbled circuit execution suggests that, with a 200 Gbps NIC, parallelism will not scale beyond $\approx$~6--7 cores.
We can solve this by installing \emph{multiple} ConnectX-7 NICs within each server.
For example, a server with 64 PCIe 5.0 lanes available for networking can theoretically support $\approx$~2~Tbit/s.
This could be scaled further by using multiple CPU sockets.

Overall, we optimistically estimate up to \emph{three orders of magnitude} performance gains in moving from a WAN to a fast LAN---one identified by prior work (top of \secref{sec:performance}), one from our work in \secref{s:garbled_circuit_experiments} and \secref{s:secret_sharing_experiments}, and one from parallelization across multiple cores and adjusting machine shape (\secref{s:future_performance_opportunities}).
This motivates \arch{} architectures that would enable \focusapps{} to benefit from fast LANs while maintaining distributed trust.
\section{Trusted Administrative Entities}
\label{sec:centralized}

In certain industry deployments, it is considered sufficient for the $n$ parties to execute on physically different servers, even if a single administrator controls all parties~\cite{sharemind} or their infrastructure~\cite{blockdaemon, alchemy}.
Certain academic works target this model too, assuming the parties use machines controlled by separate cloud accounts in the same datacenter~\cite{falk2023gigdoram}.

Our position, in this paper, is that \emph{such deployment models do not truly provide distributed trust.}
This view is echoed by systems researchers and industry practitioners working on distributed trust~\cite{dauterman2022reflections, kaviani2024flock, lindell2023mpc}.
For example, if both parties are in the same cloud datacenter, a malicious cloud employee could see both parties' memories and subvert \smpc{} entirely.
Furthermore, if all parties trust a single cloud provider, they could use a clean room~\cite{awscleanrooms, gcpcleanrooms, snowflakecleanrooms, databrickscleanrooms} instead of \smpc{}.

Still, we note that such deployments, with a mutually trusted administrator or cloud, can benefit from the research we proposed in \secref{sec:performance}.
They would just need to use optimized \smpc{} software, and server and network hardware capable of kernel bypass (e.g., RDMA).
Cloud providers would have to allow accounts controlled by \emph{different} tenants to allocate kernel-bypass VMs that are placed nearby (e.g., on the same rack).
To our knowledge, cloud providers today do not allow this, but we see no technical obstacles to them doing so.
\section{Distributed but Proximate Trust}
\label{sec:model}

How can \focusapp{} deployments benefit from careful system design for LANs, as we proposed in \secref{sec:performance}, without weakening distributed trust as in the models discussed in \secref{sec:centralized}?
In this section, we answer these questions by developing deployment models that enable parties to remain in separate trust domains while being physically near each other.
We call such architectures \emph{Distributed But Proximate Trust (\arch{}).}
Security researchers routinely measure \focusapp{} protocols over LANs~\cite{\LANWithoutExplanation}; \arch{} fills a valuable missing piece for such work, providing a basis for LAN-based deployments and grounding for LAN-based experiments.

We present our \arch{} models as a sequence of thought experiments.
First, we aim to understand if there are any fundamental roadblocks: \emph{in principle}, with sufficient resources, can $n$ parties achieve \arch{}?
We iterate on the resulting setup, moving toward designs that are more broadly accessible.

\subsection{\arch{} for Well-Resourced Parties}
\label{s:model_hyperscalars}

Let us assume that the $n$ parties who wish to run \afocusapp{} have the resources and expertise for significant investments in computing infrastructure.
This would be representative of hyperscalars, like Microsoft and Amazon, wanting to run a high-value \focusapp{}.
Are there any \emph{fundamental} obstacles that would prevent such parties from obtaining DBPT?

Our solution is for such parties to coordinate to \emph{build datacenters that are physically nearby}---for example, in different buildings in the same city block.
These need not be large facilities like traditional cloud regions, but can be small facilities similar to points of presence or edge datacenters.
This is consistent with the expansion of cloud infrastructure to the edge via smaller datacenter deployments, such as Azure Edge/Extended Zones~\cite{khalidi2020azureedgezones, azureextendedzones} and AWS Local Zones~\cite{awslocalzones}.

The datacenters' proximity enables connectivity via dedicated, high-bandwidth fiber optic cables, providing a fast network for \focusapps{}.
Whereas a global-scale WAN is extremely expensive to build and run, datacenters in the same city block can be connected cheaply via commodity hardware.
For example, a 200 Gbps fiber optic cable 150~m in length costs only $\approx$~500 USD~\cite{fiberoptic200g}, making it feasible to connect each server in one datacenter one-to-one with a corresponding server in another party's datacenter via \emph{multiple} such cables, at only a fraction of the cost of the servers themselves.
The datacenters' physical proximity would enable network latencies comparable to intra-rack networks.
For example, propagation delay over a 200~m fiber optic cable is only $\approx$~1~\us{}, comparable to processing delays within NICs~\cite{wei2023smartnic, schuh2024ccnic}.

Each party can have disjoint sets of employees manage their respective datacenters and their buildings' physical security.
Thus, from an administration standpoint, there is no entity in control of both parties' infrastructures.
While there is still some ``ambient trust,'' such as trust in hardware vendors (e.g,. Intel) and supply chains, this also applies to the status quo in which datacenters are not physically nearby.

Outside of computer administration/manufacturing, there is a minor difference in trust model: Nearby datacenters may share the same political authority, if located in the same country or municipality.
If the parties are in different countries and do not trust each other's government, then physical co-location may weaken the trust model, unless the datacenters are co-located in a way that straddles a shared national border.
That said, many applications of \focusapps{} involve parties in the same country, for which this would not be an issue.

\subsection{\arch{} via \focusapp{}-Friendly Clouds}
\label{s:model_clouds}

Not all organizations wishing to run \focusapps{} can directly use the deployment model in \secref{s:model_hyperscalars}.
For example, banks and hospitals stand to benefit from \focusapps{}, but may not have the resources or expertise to build/manage datacenters and run network cables.
How can such organizations obtain DBPT?

Our observation is that, if $c$ hyperscalars implement the deployment model in \secref{s:model_hyperscalars}, then they can make it available to other, less-well-resourced parties while preserving trust boundaries.
In effect, they can become ``cloud providers'' for a \focusapp{}-friendly cloud that other parties can use.

The idea is that each of the $n$ parties can use secret sharing to split its data into $c$ shares, and then give each provider one of the shares.
While distributing these shares requires network bandwidth, it is proportional only to the size of the data, not to the size of the computation, and therefore is likely to be much cheaper than just running \smpc{} over the public Internet.
This approach is directly inspired by prior work that separates compute providers, which act as parties in a distributed trust system, from data owners~\cite{chen2020metal, liagouris2023secrecy}.

In cryptography, this is sometimes called the ``client-server model.''
As long as the requisite number of compute providers are honest, the distributed trust guarantees of \smpc{} hold, possibly with abort.
This assumption is reasonable because the providers are well-resourced and economically motivated to not collude.
Any evidence of misbehavior would cause significant economic damage due to reputational impact, likely outweighing the value of stealing the $n$ parties' data.

\subsection{\arch{} for Everyone}
\label{s:model_colos}

The approach in \secref{s:model_clouds} has a major drawback: It requires at least two hyperscalars to invest in co-located datacenters and make them accessible to others via a cloud service.
We are not aware of any such deployment today.
How can organizations without resources or expertise to build datacenters obtain DBPT, in a way that is practical and deployable today?

Our insight is to use rentable datacenter and edge infrastructure distinct from the cloud.
Companies like Digital Realty~\cite{digitalrealtycolocation}, Equinix~\cite{equinixcolocation}, Hurricane Electric~\cite{hurricaneelectriccolocation}, Rackspace~\cite{rackspacecolocation}, etc. provide \emph{colocation services} (``colos'').
With colos, customers provide their own physical servers; the colo provides power and networking infrastructure to host their customers' servers.
Whereas cloud providers control hypervisor software running on each server, colo providers are only responsible for power and networking, and leave customers in full software control of their servers.
This provides a path to use colos for \focusapps{} without the colo provider gaining control over all parties' infrastructure.

Today, colos offer an alternative to on-premises solutions, enabling the control associated with bringing one's own servers while amortizing the operational costs (e.g., power, cooling, networking, security) of running a datacenter~\cite{equinixcolocation}.
Although it is not the typical use case of a colo, colos can be used to provide fast networks for clients whose servers are in the same building.
For example, Digital Realty provides ``Cross Connects'' that allow for dedicated network connections between different customers' physical servers~\cite{digitalrealtycrossconnects}.

A major drawback of a colo-based solution, however, is that the colo provider has physical access to both parties' computing infrastructures.
An adversarial colo provider could conduct side channel attacks to learn the parties' secret data.
A malicious colo provider could even install unauthorized hardware inside customers' servers, e.g., to snoop on the memory bus~\cite{lee2020membuster} to learn the parties' secrets.

We observe that this issue could be overcome via a new research direction on developing \emph{tamper-proof cages} for running physical servers in untrusted environments.
The idea would be to destroy any sensitive data on a server if its cage is compromised while it is running.
For example, one could install sensors in a server that detect if the cage is compromised, and if so, send a high priority interrupt to the CPU, which would repond by immediately clearing all registers and memory.
Alternatively, one could design a tamper-proof cage that simply cuts power to the device.
If an even higher degree of assurance is required, the tamper-proof cage could physically damage the server (i.e., immolation~\cite{mickens2025guillotine}).
A possible alternative to tamper-proof cages is for each organization to have an employee monitor their server in the colo environment (or hire a trusted guard company to do the monitoring).

In this paper, we leave the design and implementation of tamper-proof cages to future work, and we do not provide a recommendation as to which approach is ``best.''
We simply note that the ``physical access'' problem with using colos for \arch{} can be plausibly solved by developing such devices.
\section{Future Research and Conclusion}
\label{sec:future}

We urge the community to further develop \arch{} architectures for deploying \focusapps{} in LANs, to realize the performance gains in \secref{sec:performance}.
In particular, the \arch{} model in \secref{s:model_colos} has the potential to work broadly, as it does not require the buy-in of hyperscalars.
Research on tamper-proof cages can help enable this model, and thereby \arch{}.

\arch{} enables additional performance gains, beyond those in \secref{sec:performance}, because it shifts the balance between CPU and networking.
Because \arch{} makes high-bandwidth and low-latency communication cheaply available to \focusapps{}, one could potentially improve \focusapp{} performance by reducing CPU usage at the expense of higher network usage (to an extent).
For example, \focusapp{} systems can adjust how they combine bandwidth-intensive garbled-circuit-based \smpc{} and round-intensive secret-sharing-based \smpc{} to strike a new balance.
On the theory side, there is an opportunity to develop new distributed trust protocols, including for distributed ORAM~\cite{vadapalli2023duoram, doerner2017floram, falk2023gigdoram}, LPZK~\cite{dittmer2021lpzk, dittmer2022improving}, PSI~\cite{freedman2004psi, huang2012psi}, secret-sharing-based \smpc{}~\cite{benor1988bgw}, etc., to find different trade-offs between CPU and networking.
For example, a major focus in garbled-circuit-based \smpc{} design over the past two decades has been to reduce the amount of data transferred between the parties~\cite{kolesnikov2008freexor, rosulek2015history, zahur2015halfgates, yakoubov2017gentle, liu2025bitgc, liu2025authenticatedbitgc}; \arch{} suggests making trade-offs in the opposite direction.

\bibliographystyle{ACM-Reference-Format} 
\bibliography{db}

\end{document}